\documentstyle[pra,aps,twocolumn,epsfig]{revtex}
\begin{document}
\title{High efficiency Nondistortion Quantum Interrogation of atoms in quantum superpositions}
\author{Xingxiang Zhou$^1$, Zheng-Wei Zhou$^2$,
     Guang-Can Guo$^2$, Marc J. Feldman$^1$}

\address{
$^1$Superconducting Digital Electronics %and Quantum Computing
Lab, Electrical and Computer Engineering Department, \\ University
of Rochester, Rochester, NY 14627, USA \\ $^2$Laboratory of
Quantum Communication and Quantum Computation and Department of
Physics,\\ University of Science and Technology of China, Hefei,
Anhui 230026, China \\}
\date{\today}
\maketitle
%\narrowtext

\begin{abstract}
We consider the nondistortion quantum interrogation (NQI) of an
atom prepared in a quantum superposition. By manipulating the
polarization of the probe photon and making connections to
interaction free measurements of opaque objects, we show that
nondistortion interrogation of an atom in a quantum superposition
can be done with efficiency approaching unity. However, if any
component of the atom's superposition is completely transparent to
the probe wave function, a nondistortion interrogation of the atom
is impossible.
\end{abstract}


Interaction-free measurements (IFM) were first considered by
Elitzur and Vaidman to illustrate the peculiar nonlocality of
quantum mechanics \cite{ref:EV93}. %In their original argument, a
%photon passes through a Mach-Zehnder interferometer arranged in a
%way such that the photon exits from one port (the bright port)
%with certainty. %When there is an absorbing object, in their
%original proposal an ultra-sensitive ``exploding bomb'', in one of
%the interferometer paths, the interference of the photon wave
%function is modified and
%When an absorbing object, in their original proposal an
%ultra-sensitive ``exploding bomb'', blocks the photon wave
%function in one of the interferometer paths, the interference of
%the photon wave function is modified and there is a non-vanishing
%probability that the photon will exit from the dark port. In this
%way the presence of the absorbing object can be inferred without
%the probing photon actually being absorbed, hence the name
%``interaction-free measurement''.
It was shown that it is possible to infer the presence of an
absorbing object (in their original argument an ultra-sensitive
``exploding bomb'') in a Mach-Zehnder interferometer without the
probe photon being absorbed by the object. This works because the
absorbing object blocks any photon passing it and changes the
interference of the photon wave function. Since the original
proposal of IFM, there have been many theoretical and experimental
studies on this issue. It was shown that interaction-free
measurements can in principle be done with unit efficiency in an
asymptotic sense, for both opaque
\cite{ref:Kwiat95},\cite{ref:Kwiat99} and semi-transparent objects
\cite{ref:semi1},\cite{ref:semi}.

As emphasized by Vaidman \cite{ref:free?}, IFM's are not
necessarily initial state preserving measurements. Due to the
non-vanishing interaction Hamiltonian, in general IFM's can change
very significantly the quantum state of the object being observed.
However in most cases we wish to do the IFM without changing the
internal state of the observed object, which we may call a
``nondistortion quantum interrogation'' (NQI) \cite{ref:ours}. In
most previous treatments, it was claimed that interaction-free
measurements can be done for a quantum mechanical object as well
as for the ``exploding bomb'' discussed in the original proposal
\cite{ref:EV93}. For a two-level atom in its ground state
interacting with a resonant photon, this is certainly true since
the absorption of the photon destroys the initial state of the
atom completely. However, the claim that IFM can be done equally
well for a quantum mechanical object as for an ``exploding bomb''
is not fully justified unless the quantum superposition of the
quantum object is taken into account. After all, the possibility
of being in distinct states simultaneously is what distinguishes
quantum from classical \cite{ref:von}. As discussed in a recent
paper by P\"{o}tting {\em et. al} \cite{ref:main}, the IFM and NQI
of an atom in quantum superposition are more subtle than those of
a classical object since the atom is subject to measurement
dependent decoherence. Though in general NQI schemes can be
designed for an atom in a quantum superposition \cite{ref:ours},
\cite{ref:main}, the previous schemes based on a simple
Mach-Zehnder interferometer setup yield very low success
probabilities.

%\begin{figure}[h]
%    \centering
%    \epsfig{file=atom.eps, width=2in, height=1.8in}
%    \caption{Level structure of the atom. The atom can make a transition
%    to the excited state $|e\rangle$ from $|m_+\rangle$ or $|m_-\rangle$ by
%    absorbing a circular polarized photon. It then decays rapidly to the
%    stable ground state $|g\rangle$.} \label{fig:atom}
%\end{figure}

In this work we show that nondistortion interrogation of an atom
in quantum superposition can be done with efficiency approaching
unity, by using the model of \cite{ref:main} and making
connections to IFM's of opaque objects. However, a necessary
condition for such an NQI is that the possibility of interaction
exists between the probe and every component of the superposition.
It is then easily proved that an NQI of the atom in a quantum
superposition is impossible if any component of the superposition
is completely transparent to the probe.
\begin{figure}[h]
    \centering
    \epsfig{file=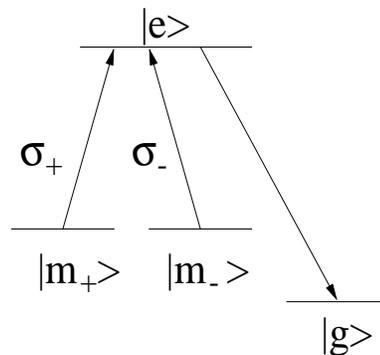, width=2in, height=1.8in}
    \caption{Level structure of the atom. The atom can make a transition
    to the excited state $|e\rangle$ from $|m_+\rangle$ or $|m_-\rangle$ by
    absorbing a circularly polarized photon. It then decays rapidly and
    irreversibly to the stable ground state $|g\rangle$.} \label{fig:atom}
\end{figure}

As in Fig. \ref{fig:atom}, the model we consider is a multilevel
atom prepared in a superposition of the two degenerate metastable
states $|m_+\rangle$ and $|m_-\rangle$. Starting from
$|m_+\rangle$ and $|m_-\rangle$, the atom can absorb a $+$ or $-$
(circularly) polarized photon and make a transition to the excited
state $|e\rangle$ with unit efficiency. It then decays
irreversibly to the ground state $|g\rangle$ very rapidly. The
whole process is
\begin{equation}
|\pm\rangle|m_\pm\rangle\longrightarrow |S_\pm\rangle|g\rangle
\label{eq:abs}
\end{equation}
where $|\pm\rangle$ are the $+$ or $-$ polarized incident photons
and $|S_\pm\rangle$ are the corresponding scattered photons which
we assume will not be re-absorbed by the atom and can be filtered
away from the detectors. The state of the atom is in the
superposition
\begin{equation}
|\psi_{atom}\rangle=\alpha|m_+\rangle+\beta|m_-\rangle
\label{eq:state}
\end{equation}
where $\alpha$ and $\beta$ are unknown non-vanishing amplitudes
satisfying $|\alpha|^2+|\beta|^2=1$.

If we can use a photon that will be completely absorbed by the
atom, then the problem is identical to that of an opaque object.
However, no matter what the photon's polarization is ($+$ or $-$
or a superposition of them), it will only be partially absorbed by
the atom, due to the polarization selective interaction
(\ref{eq:abs}). For instance, the direct interaction between an
$x$ polarized photon $\frac{1}{\sqrt{2}}(|-\rangle-|+\rangle)$ and
the atom results in the state
\begin{equation}
\frac{1}{\sqrt{2}}(\alpha|-\rangle|m_+\rangle-
\beta|+\rangle|m_-\rangle)-\frac{1}{\sqrt{2}}(\alpha|S_+\rangle-\beta|S_-\rangle)|g\rangle
\end{equation}
If the probe photon is not actually scattered, the photon and atom
end up in the entangled state
$\alpha|-\rangle|m_+\rangle-\beta|+\rangle|m_-\rangle $. As shown
in \cite{ref:main}, if we do not change the polarization of the
photon through the interrogation process, this partial absorption
and entangling will change the state of the atom even if no
absorption happens, and result in a very low efficiency for the
NQI of the atom.

At this point it might seem that an NQI of the atom in quantum
superposition is similar to that of semi-transparent objects
\cite{ref:semi}, since no complete absorption could happen if we
do not do anything on the polarization of the photon. This is not
true though. Once the wave functions of the photon and atom are
entangled, the atom becomes transparent to the photon and it will
not interact with the photon again when the photon passes it a
second time. On the other hand, we can make a connection to NQI's
of both opaque and semi-transparent objects if we let the photon
pass the atom twice, with its polarization changed from the
original value the second time. For instance, if we use a $+$
polarized photon to interact (directly) with the atom prepared in
(\ref{eq:state}), we end up in the state
$\beta|+\rangle|m_-\rangle+\alpha|S_+\rangle|g\rangle$ the first
time. If no absorption actually happens, the photon and atom are
left in $|+\rangle|m_-\rangle$. We then change the polarization of
the photon to $-$ and let it pass the atom a second time. This
time the photon will be absorbed by the atom with certainty. In
this way the atom in superposition is effectively an opaque object
to the photon. In the following, we show two ways of
unit-efficiency (in an asymptotic sense) NQI of the atom in a
superposition, following this idea of polarization rotation.

\begin{figure}[h]
    \centering
    \epsfig{file=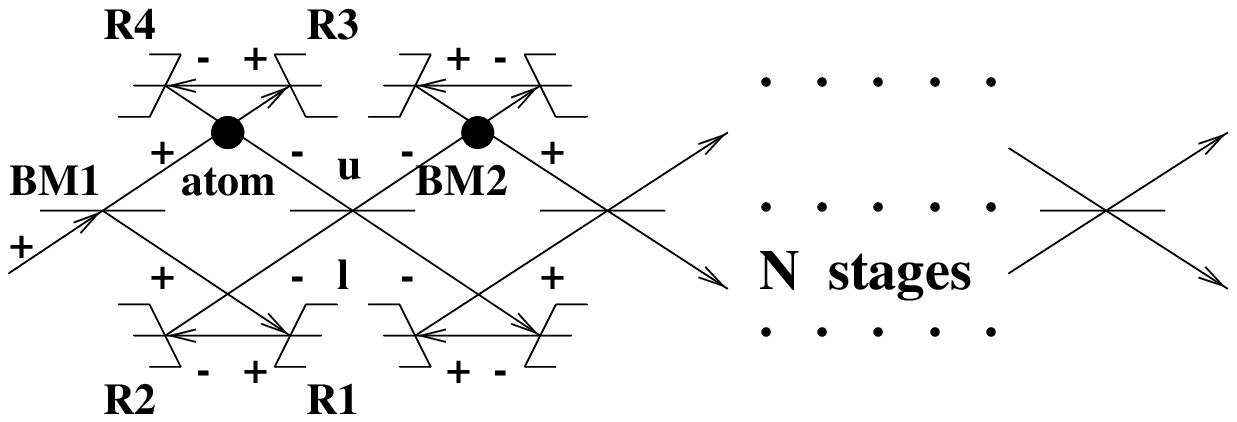, width=3.4in, height=1.6in}
    \caption{Nondistortion interrogation of the atom with a series of
Mach-Zehnder interferometers. Between $R1$ and $R2$, $R3$ and
$R4$, the polarization of the photon is rotated to the orthogonal
value. In absence of the atom, for properly chosen reflectivity of
the beam splitters the photon will exit through the upper port of
the last interferometer with certainty. When the atom is present,
any photon going into the upper half of the interferometer is
absorbed after it passes the atom twice. %So the
%interference of the photon wave function is modified and
In this case there is a finite probability that the photon exits
through the lower port
of the last interferometer. %This probability goes to unity when
%the reflectivity of the beam splitters approaches unity and the
%number of stages goes to infinity.
}\label{fig:MZ}
\end{figure}

In Fig. \ref{fig:MZ} we consider the folded Mach-Zehnder
interferometer discussed in \cite{ref:Kwiat95}. For the purpose of
clarity it is drawn in the form of $N$ Mach-Zehnder
interferometers connected in series, therefore the atom is in
every single interferometer (the dot). Each interferometer
consists of two beam splitters ($BM1$ and $BM2$) and four
reflecting mirrors ($R1$, $R2$ and $R3$, $R4$). $R3$ and $R4$ are
used to redirect the photon to the atom after it passes the atom
the first time. Suppose the probe is a $+$ polarized photon
incident from the lower left port to the first interferometer. The
reflectivity of each beam splitter is $R=\cos^2(\pi/2N)$ and the
phase difference between the upper and lower paths is zero. In
addition, the polarization of the photon is rotated to the
orthogonal one (from $+$ to $-$ or from $-$ to $+$) when the
photon travels between $R1$ and $R2$, $R3$ and $R4$ (There are
many ways to do this, for instance by using a half wave plate). At
$BM2$ the upper and lower branches of the photon wave function are
in the same polarization (even though the polarization is
orthogonal to that of the incident photon), so the interference
between them is maintained. In absence of the atom, after $N$
stages the photon will exit with certainty from the upper port of
the last interferometer, with its polarization unchanged if $N$ is
even or rotated to the orthogonal value if $N$ is odd.

Now we see that the interference of the photon wave function is
changed completely if the atom is in the interferometers (assume
it is in the upper half of the system). Starting from the incident
point, let us trace the wave function of the system (photon plus
atom) until the photon arrives at $BM2$:
\begin{eqnarray}
\lefteqn{|+\rangle_l(\alpha|m_+\rangle+\beta|m_-\rangle)
\stackrel{BM1}{\longrightarrow}
(t|+\rangle_u+ir|+\rangle_l)(\alpha|m_+\rangle+\beta|m_-\rangle)}
\nonumber \\ & &\stackrel{atom}{\longrightarrow} \alpha
t|S_+\rangle|g\rangle+\beta t
|+\rangle_u|m_-\rangle+ir|+\rangle_l(\alpha|m_+\rangle+\beta|m_-\rangle)
\nonumber \\ & &\stackrel{R's}{\longrightarrow} \alpha
t|S_+\rangle|g\rangle-\beta t
|-\rangle_u|m_-\rangle-ir|-\rangle_l(\alpha|m_+\rangle+\beta|m_-\rangle)
\nonumber \\ & &\stackrel{atom}{\longrightarrow} t(\alpha
|S_+\rangle-\beta
|S_-\rangle)|g\rangle-ir|-\rangle_l(\alpha|m_+\rangle+\beta|m_-\rangle)
\end{eqnarray}
where $l$ and $u$ denote the lower and upper path and
$(t,r)=(\sin(\pi/2N),\cos(\pi/2N))$ are the amplitude transmission
and reflection coefficients of the beam splitters. We have
neglected the phase advance of the photon wave function in the
above, since it is the same for the upper and lower branches (we
assumed the photon wave function picks up a phase shift of $i$
each time it is reflected). As expected, any photon that enters
the upper half of the system is absorbed. On the other hand the
reflected photon wave function is in a direct product with the
initial state of the atom. So this is equivalent to the NQI of an
opaque object, and after $N$ stages the probability that the
photon exits through the lower port (thus a successful NQI of the
atom) is
\begin{equation}
P_{NQI}=[\cos^2(\pi/2N)]^N
\end{equation}
which in the limit of large $N$ goes to unity. As pointed out in
\cite{ref:main}, this way of unit-efficiency NQI can be viewed as
a discrete form of the quantum Zeno effect \cite{ref:zeno}.

\begin{figure}[h]
    \centering
    \epsfig{file=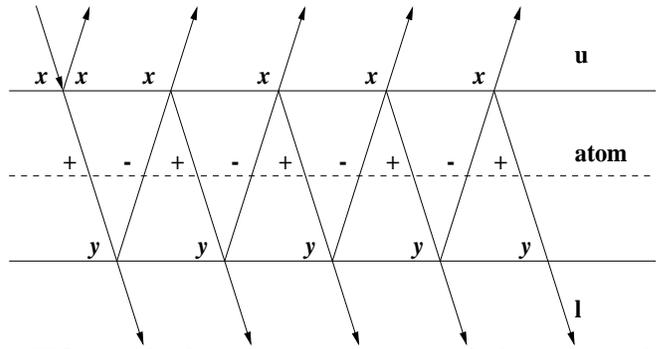, width=3.4in, height=1.8in}
    \caption{Nondistortion interrogation of the atom with
    Febry-Perot interferometer. Polarization of the photon is
    rotated from $x$ to $+$, $y$, $-$ and back to $x$ when the
    photon is reflected once in the interferometer.
    }\label{fig:FP}
\end{figure}

Fabry-Perot interferometer can also be used to do NQI's of the
atom \cite{ref:Tsegaye98}, \cite{ref:semi}. In Fig. \ref{fig:FP},
the incident photon is linearly ($x$) polarized. (The photon is
assumed to be normally incident but for clarity it is depicted as
if the angle of incidence was nonzero).
%($\cos(-\pi/4)|+\rangle+\sin(-\pi/4)|-\rangle$).
In the Fabry-Perot interferometer, its polarization changes in the
following way: when it goes through the upper half of the
Fabry-Perot interferometer, its polarization is changed to $+$.
The polarization is rotated to $y$ when the photon goes though the
lower half of the interferometer. When it is reflected back, its
polarization is changed to $-$ and back to $x$. This can be done
for instance by using a properly oriented half wave plate in the
interferometer. So all the reflected and transmitted beams are in
$x$ and $y$ polarization respectively. Assume the phase difference
between adjacent reflected or transmitted beams is $4\pi$ (so all
reflected and transmitted beams are in phase). Suppose the
possible location of the atom is in the middle of the
interferometer (represented by the dashed line). It is easily seen
that when no atom is in the interferometer the interference of the
reflected and transmitted beams is such that the photon goes
though the interferometer with certainty, for any values of the
amplitude reflection and transmission coefficients. In presence of
the atom, in exactly the same way as described before the photon
wave function that goes into the interferometer gets completely
absorbed by the atom after it passes the atom twice. The final
state of the photon-atom system is
\begin{equation}
ir|x\rangle_r(\alpha|m_+\rangle+\beta|m_-\rangle)+tt'\beta|y\rangle_t|m_-\rangle
+|abs\rangle
%+t(r\beta|S_-\rangle-\alpha|S_+\rangle)|g\rangle
\end{equation}
where $|x\rangle_r$ and $|y\rangle_t$ are the reflected and
transmitted photons (in $x$ and $y$ polarization) and
$|abs\rangle$ (unnormalized) corresponds to the situation that the
photon is absorbed. $r$ is the amplitude reflection coefficient
when the photon goes into the interferometer, $t$ and $t'$ are the
amplitude transmission coefficients when the photon goes into and
out of the interferometer. When the photon is reflected, the
superposition of the atom is unperturbed and a successful NQI is
realized. The probability of a successful NQI is $|r|^2$, which
goes to unity when $|r|\rightarrow1$.

In the above we showed that indeed high efficiency NQI's for atoms
in quantum superpositions can be realized, through the connection
to opaque objects. If we go beyond the model shown in Fig.
\ref{fig:atom} and consider other situations, for instance
%an atom in superposition of ground and excited states interacting
%with a resonant photon or
a system similar to that of Fig. \ref{fig:atom} but with
non-degenerate $|m_+\rangle$ and $|m_-\rangle$, what is the
restriction in the more general cases? In the case of high
efficiency IFM's for opaque objects, Kwiat {\em et. al} pointed
out that in order to reduce the probability that an interaction
occurs it is crucial that the {\em possibility} of such an
interaction exists \cite{ref:Kwiat99}. In the following we prove
that the necessary condition for a successful NQI of an atom in a
quantum superposition is that the possibility of interaction
exists between the probe wave function and every component of the
superposition.

%We need only consider the case that the atom is in a superposition
%of two (orthogonal) states. This is because the possibility for
%NQI's of a two dimensional system is a necessary condition for
%that of a $N(\geq 2)$ dimensional system, since one special case
%of the superposition in an $N$ dimensional system is that all
%coefficients vanish except for 2 states.
We prove this by making use of a general formalism by Mitchson and
Massar \cite{ref:semi}, with the additional requirement that the
initial state of the atom must be kept unchanged. Suppose that the
Hilbert space of the atom is an $N$($\geq 2$) dimensional space
spanned by the orthonormal base vectors $\{|\Psi_{a,j}\rangle,
j=1,2...N\}$. The NQI starts with
$|\Psi_P^i\rangle|\Psi_a^i\rangle$, where $|\Psi_P^i\rangle$ and
$|\Psi_a^i\rangle$ are initial states of the probe and atom
respectively. The atom is prepared in the arbitrary and unknown
superposition state
\begin{equation}
%|\Psi_a^i\rangle=a_1|\Psi_{a,1}\rangle+a_2|\Psi_{a,2}\rangle
|\Psi_a^i\rangle=\sum_{j=1}^{M}a_j|\Psi_{a,j}\rangle
\label{eq:sup}
\end{equation}
where $a_j's$ are unknown non-vanishing coefficients and $2\leq M
\leq N$. In the process of the interrogation, there are several
steps in which the probe and atom are arranged in such a way that
an interaction can potentially occur (the so called ``I steps'' in
\cite{ref:semi}). In between these steps unitary operations are
performed on the probe wave function. The NQI fails and stops when
an interaction between the probe and atom actually happens. If
this is not the case the state of the probe is measured at the
end. (A protocol in which the probe is measured before the end can
be converted to this form \cite{ref:semi}, \cite{ref:equi}). First
consider the case that the atom is in the non-superposed state
$|\Psi_{a,l}\rangle$ (all other components vanish). If the atom is
not in the interferometer, no interaction between the photon and
atom could occur, and the effect of the NQI before the final
measurement is an overall unitary operation on the probe:
$|\Psi_P^i\rangle|\Psi_{a,l}\rangle\rightarrow
|\Psi_{P,a,l}^f\rangle=U_P|\Psi_P^i\rangle|\Psi_{a,l}\rangle
=|\Psi_P^f\rangle|\Psi_{a,l}\rangle$, where
$|\Psi_P^f\rangle=U_P|\Psi_P^i\rangle$ and $l=1,2...N$. In
presence of the atom, the interaction could happen, but the state
of the atom will not be affected if the interaction does not
actually happen (note this is only true for $|\Psi_{a,l}\rangle's$
but not for a superposition of them). So the final state is
$|\Psi_{P,a,l}^{f'}\rangle=|\Psi_{P,l}^{f'}\rangle|\Psi_{a,l}\rangle+|interacted\rangle$
instead ($|\Psi_{P,l}^{f'}\rangle$ and $|interacted\rangle$
unnormalized), where $|\Psi_{P,l}^{f'}\rangle\neq
|\Psi_P^{f}\rangle$ in general and $|interacted\rangle$
corresponds to the situation that an interaction happens. When the
atom is in the superposition (\ref{eq:sup}), the final state in
absence of the atom is
\begin{equation}
|\Psi_{P,a}^f\rangle=|\Psi_P^f\rangle|\Psi_a^i\rangle=
|\Psi_P^f\rangle\sum_{j=1}^Ma_j|\Psi_{a,j}\rangle%+a_2|\Psi_{a,2}\rangle)
\label{eq:absf}
\end{equation}
On the other hand when the atom is present the final state is
\begin{equation}
|\Psi_{P,a}^{f'}\rangle=\sum_{j=1}^Ma_j|\Psi_{P,a,j}^{f'}\rangle
%+a_2|\Psi_{P,a,2}^{f'}\rangle
\label{eq:pref}
\end{equation}
%For the presence of the atom to be discovered, the probe must end
%up in some state different from $|\Psi_P^f\rangle$.% (the probe and
%atom could be entangled).
% In addition, the NQI requires that the initial state of the atom
%be kept when the final state of the probe is measured in some
%state orthogonal to $|\Psi_P^f\rangle$.
We can see that the necessary condition that a successful NQI can
be done is that there exists a projector
$P=|\Phi_P\rangle\langle\Phi_P|\otimes I_a$ which satisfies
$P|\Psi_{P,a}^f\rangle=0$ and
$P|\Psi_{P,a}^{f'}\rangle=\Delta|\Phi_P\rangle|\Psi_a^i\rangle$,
where $|\Phi_P\rangle\neq 0$ is some state of the probe orthogonal
to $|\Psi_P^f\rangle$, $I_a$ is the unity operator for the atom
and $\Delta$ is some nonzero number \cite{ref:necce}. %This means
%that the probe can be measured to be in a state orthogonal to that
%corresponding to the absence of the atom, which reveals the atom's
%existence.
This is because an NQI requires that the probe can be measured in
some final state orthogonal to $|\Psi_{P}^f\rangle$ (which reveals
the atom's presence) with the atom's initial state
unchanged. %in absence of the atom the probe will never be measured
%in $|\Phi_P\rangle$, while a measurement with $P$ will reveal the
%presence of the atom with its state unchanged if the above
%conditions are satisfied.
Now assume that some of the $M$ components in (\ref{eq:sup}), say
$|\Psi_{a,i}\rangle, i=1,2...K (K\leq M)$ are completely
transparent to the probe,
%interaction between $|\Psi_P^i\rangle$ and
%$|\Psi_{a1}^i\rangle$ is possible while $|\Psi_{a2}^i\rangle$ is
%completely transparent to $|\Psi_P^i\rangle$,
either due to a vanishing interaction Hamiltonian between them or
the design of the protocol. Then through the interrogation process
the wave function of the system evolves as follows:
\begin{eqnarray}
|\Psi_P^i\rangle\sum_{j=1}^M a_j|\Psi_{a,j}\rangle\rightarrow
|\Psi_{P,a}^{f'}\rangle=|\Psi_P^f\rangle\sum_{j=1}^K
a_j|\Psi_{a,j}\rangle \nonumber \\+\sum_{j=K+1}^M
a_j|\Psi_{P,a,j}^{f'}\rangle
%|\Psi_P^i\rangle(a_1|\Psi_{a,1}\rangle+a_2|\Psi_{a,2}\rangle)
%=a_1|\Psi_P^i\rangle|\Psi_{a,1}\rangle+a_2|\Psi_P^i\rangle|\Psi_{a,2}\rangle
%\nonumber \\ \rightarrow
%|\Psi_{P,a}^{f'}\rangle=a_1|\Psi_{P,a,1}^{f'}\rangle
%+a_2|\Psi_P^f\rangle|\Psi_{a,2}\rangle
\end{eqnarray}
Suppose the projector P for a successful NQI exists, the operation
with $P$ on $|\Psi_{P,a}^{f'}\rangle$ results in:
\begin{eqnarray}
P|\Psi_{P,a}^{f'}\rangle=\sum_{j=K+1}^M
a_jP|\Psi_{P,a,j}^{f'}\rangle \nonumber \\
=\sum_{j=K+1}^M a_j\langle\Phi_P
|\Psi_{P,j}^{f'}\rangle|\Phi_P\rangle|\Psi_{a,j}\rangle
\label{eq:proj}
\end{eqnarray}
Obviously, an NQI in this case is impossible, since the right hand
side of (\ref{eq:proj}) does not contain any $|\Psi_{a,l}\rangle,
l=1,2...K$ component. This is easy to understand, because the
non-interaction between the probe and $|\Psi_{a,l}\rangle
(l=1,2...K)$ makes it impossible to change the evolution of that
branch of the wave function ($|\Psi_P^i\rangle
|\Psi_{a,l}\rangle$). When the final state of the probe is
measured using a projector orthogonal to
$|\Psi_P^f\rangle\langle\Psi_P^f|$, all components that are
completely transparent to the probe drop out of the atomic wave
function.

This result explains why an NQI of the atom in superposition is
impossible if a $+$ or $-$ photon is used as the probe and nothing
is done on its polarization \cite{ref:main}. On the other hand, an
NQI with linearly polarized probe photon is possible. (Actually
our high-efficiency schemes work with linearly polarized photon
too if its polarization is manipulated the same way we
prescribed.) Also, an NQI for a system similar to that in Fig.
\ref{fig:atom} but with nondegenerate $|m_+\rangle$ and
$|m_-\rangle$ is impossible if one uses a single probe photon in
resonance with one (but not both) of the two metastable states
$|m_+\rangle$, $|m_-\rangle$.

In summary, we showed that an nondistortion interrogation of an
atom in a quantum superposition can be done with efficiency
approaching unity, by making the photon wave function interact
with all components of the superposition and turning the problem
to that of an opaque object. On the other hand if any component of
the superposition is transparent to the probe wave function, such
an NQI is impossible.

Work of X. Zhou and M. J. Feldman was supported in part by ARO
grant DAAG55-98-1-0367. Z. W. Zhou and G. C. Guo are funded by
National Natural Science Foundation of China.

\end{document}